# Doping of zigzag carbon nanotubes through the encapsulation of small fullerenes

**K. S. Troche[1], V. R. Coluci[1]\*, R. Rurali[2], and D. S. Galvão[1]**

**[1]Instituto de Física Gleb Wataghin, Universidade Estadual de Campinas
CP 6165, CEP 13083-970, Campinas-SP, Brazil
[2]Departament d'Enginyeria Electrònica,
Escola Tècnica Superior d'Enginyeria (ETSE),
Universitat Autònoma de Barcelona (UAB)
08193 Bellaterra (Barcelona), Spain**

## Abstract

In this work we investigated the encapsulation of $C_{20}$ and $C_{30}$ fullerenes into semiconducting carbon nanotubes to study the possibility of bandgap engineering in such systems. Classical molecular dynamics simulations coupled to tight-binding calculations were used to determine the conformational and electronic properties of carbon nanotube supercells containing up to 12 fullerenes. We have observed that $C_{20}$ fullerenes behave similarly to a p-type dopant while $C_{30}$ ones work as n-type ones. For larger diameter nanotubes, where fullerene patterns start to differ from the linear arrangements (peapods), the doping features are preserved for both fullerenes, but local disorder plays an important role and significantly alters the electronic structure. The combined incorporation of both fullerene types (hybrid encapsulation) into the same nanotube leads to a behavior similar to that found in electronic junctions in Silicon-based electronic devices. These aspects can be exploited in the design of nanoelectronic devices using semiconducting carbon nanotubes.

- **Corresponding author: coluci@ifi.unicamp.br**

**Introduction**

The discovery of fullerenes[1] and carbon nanotubes[2] opened a new and enormous field in theoretical and experimental research of these materials. The development and improvement in the manipulation and fabrication methods of these new structures, specially carbon nanotubes (CNTs), have lead to a very large number of theoretical and experimental studies for these materials with the necessity of the understanding and interpretation of the obtained data, as well as their exploration of promising properties due to its rich electronic and/or mechanical characteristics.

Depending upon structure CNTs are either metallic or semiconducting, which is a feature that has been intensively investigated and exploited in prototype devices. The metallic characteristic is presented by armchair ($n,n$) CNTs while semiconducting features are observed in zigzag ($n,0$) ones, when $n$ is not multiple of 3. However, for zigzag tubes with large diameters, i.e., for large values of $n$, the metallic behavior is again observed [3].

A very interesting research area is the one related to the tunning of the electronic and mechanical properties through the process of incorporation of organic and inorganic compounds into CNTs. In this way, the encapsulation of $C_{60}$ into CNTs, generically called peapods [4], provides in principle possibilities to investigate a large class of phenomena in chemistry and physics, such as well formed ordered molecular phases induced by packing effects [5-6], and the transformation of peapods into double-walled carbon nanotubes under thermal process [7].

The encapsulation of $C_{60}$ within single-walled CNTs changes the vibrational modes associated with the expansion and contraction of the tubes [8]. The electronic structure is also altered by the encapsulation. Okada et al.[9], using density functional theory within the local density approximation, have observed that (8,8), (9,9), and (10,10) peapods are metallic but, while the (8,8) and (9,9) peapods have the charge density at the Fermi level distributed along the walls of the tube, the (10,10) peapod shows a distribution on the wall as well as on the $C_{60}$ chain. This is by the endothermic character of the encapsulation of $C_{60}$ within the (8,8) and (9,9) which leads to a substantial nanotube structural distortion, producing that the peapod energy bands are not simply the sum of the constituent parts.

For semiconducting tubes, Esfarjani *et al.* [10] have proposed that using donor atoms on one side and acceptors on the other, (10,0) CNTs can work as nano diodes, showing a nonlinear rectifying effect. Lee et al. [11] have also shown that metallofullerenes can be used as band-gap modulators in semiconducting CNTs and suggested that with the insertion of different types of metallofullerenes, one-dimensional heterostructures can be produced and complex band-gap engineering could be possible [12].

While the encapsulation of $C_{60}$ in small diameter single-walled CNTs occurs as an endothermic process [9], other smaller fullerenes such as $C_{20}$ and $C_{30}$ could be used as candidates to modify properties of semiconducting CNTs through the peapod configuration. In the last years the interest for fullerenes smaller than $C_{60}$ have increased after the chemical synthesis of the cage $C_{20}$ [13] and the studies about the amazing thermodynamical stability of the fullerenes $C_i$ ($i$ = 30-82)[14-15]. Recently, Zhou *et al*. [16] have theoretically investigated three-dimensional configurations of encapsulated $C_{20}$ and $C_{28}$ in single walled carbon nanotubes.

In this work we investigated the encapsulation of $C_{20}$ and $C_{30}$ in small diameter zig-zag semiconducting carbon nanotubes to study the possibility of band-gap engineering in such systems.

**Methodology**

We have considered three zig-zag ($n$,0) ($n$ =11, 13 and 17) single-walled carbon nanotubes as the confining structures for the fullerenes. Supercells made of 30 unit cells of the zig-zag nanotubes were constructed corresponding approximately to a length of 126 Å. Periodic boundary conditions were applied along the axial direction. For the other directions the tube was placed 60 Å apart from the others in order to avoid interaction with the periodic images of the neighboring cells.

For each zig-zag nanotube, 1 up to 12 fullerenes $C_i$ ($i$ = 20 or 30) were inserted into the tubes. The first (second) group of the structures corresponds to molecules named $k$$C_{20}$@($n$,0) ($k$$C_{30}$@($n$,0)) with $n$=11, 13, or 17, and $1 \leq k \leq 12$. A third group was composed by the SWNTs filled with 50% of $C_{20}$ and 50% of $C_{30}$, and named $l$$C_{20}$$C_{30}$@($n$,0) with $n$ =11, 13, or 17, $1 \leq l \leq 4$ corresponding to the number of fullerenes $C_{20}$ and $C_{30}$ encapsulated into the nanotube. The total number of atoms for the different peapod configurations varied from 1300 up to 2100 carbon atoms.

The supercells of CNTs were initially optimized using the universal force field [17-18], implemented in the Cerius2 package [19]. This force field includes van der Waals, bond stretch, bond angle bend, inversion, torsion and rotation terms and has been used with success in the study of dynamical properties of crystalline and complex carbon nanostructures [20]. In this work we have used the Lennard-Jones 6-12 potential to describe the van der Waals interactions between carbon atoms,

$$E_{vdW} = D_{ij} \left\{ -2 \left[ \frac{x_{ij}}{x} \right]^6 + \left[ \frac{x_{ij}}{x} \right]^{12} \right\} , (1)$$

where $D_{ij}$ equals to 0.105 kcal/mol and $x_{ij}$ is equals to 3.851 Å.

We carried out classical molecular dynamics simulations of $C_{20}$ and $C_{30}$ molecules into the CNTs in the following way:

- Firstly, the fullerenes were placed near the edges of the nanotube.
- Secondly, a NVT molecular dynamics simulation at 300 K was run during hundreds of ps (time step =1 fs) in order to encapsulate the fullerene leading to the structure $kC_i@(n,0)$. The formed structure was then geometrically optimized to obtain the minimum energy conformation of the system.
- This procedure was repeated until $k$=12.

The CNTs were kept fixed during the encapsulation procedure. This mimics the encapsulation in multi-walled carbon nanotubes [21] since the deformation of the inner tube is expected to be smaller when compared with the one for the corresponding single-walled carbon nanotube. Furthermore, due to its distance to the center of the double-walled CNT, the outmost tube does not affect the arrangement of the fullerenes in the most internal tube.

With the obtained geometries from the molecular dynamics simulations the electronic band structure of $kC_i@(n,0)$ and $lC_{20}C_{30}@(n,0)$ were calculated using the tight binding model by Porezag [22] implemented in the TROCADERO program[23]. This model employs a non-orthogonality s-p basis, in which the hopping matrix elements are obtained directly from density functional theory calculations using the same basis set but disregarding three-center contributions to the Hamiltonian. This method has been successfully applied to the prediction of allotropic forms of carbon [24] and has been proved to combine accuracy and reduced computational effort, especially for large systems. The use of the γ(gamma)-point for the Brillouin-zone sampling was sufficient for the total energy convergence for the structures considered here.

**Results and Discussion**

Depending on the relation between the nanotube and fullerene diameters (and assuming an absence of dynamical barriers) the encapsulation process can be either endo ($\Delta E > 0$) or exothermic ($\Delta E < 0$) (equation (2)).

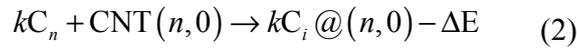

$$kC_n + \mathrm{CNT}(n,0) \rightarrow kC_i @ (n,0) - \Delta E \qquad (2)$$

Previous theoretical studies have demonstrated that the incorporation of $C_{20}$ into (13,0) nanotube is an exothermic process [9]. In order to obtain information about the encapsulation process of the structures investigated here, we estimated the required energy for the incorporation of one single molecule of $C_{20}$ ($C_{30}$) inside the CNT ($n$,0) for $n$ = 11, 13 and 17, by calculating the total energy of the configurations represented in equation (2).

We have obtained (Table 1) that the incorporation of $C_{20}$ and $C_{30}$ fullerenes into the CNT (11,0) is an endothermic process, but it is an exothermic one to the cases of CNTs (13,0) and (17,0).

| Carbon Nanotube | $\Delta E$ (Kcal/mol) | |
|:---:|:---:|:---:|
| | $C_{20}$ | $C_{30}$ |
| (11,0) | 156 | 432 |
| (13,0) | -62 | -26 |
| (17,0) | -63 | -32 |

**Table 1**: Formation energy for the encapsulation of one single fullerene into the zigzag nanotubes considered in this work. Results from TROCADERO calculations.

We have analyzed the electronic structure of the nanotubes containing $C_{20}$ and $C_{30}$ for two different configurarions: Firstly, we have considered the case of the "pure" encapsulation, i.e., only one type of fullerene inside the nanotube. Secondly, the cases of a ordered "hybrid" encapsulation (both types present, $C_{20}$s at one side of the CNT and $C_{30}$s at the opposite side) were also considered.

The encapsulation patterns of $C_{20}$ into the (11,0) and (13,0) CNTs form linear arrangements, while for the (17,0) CNT the structure of the fullerene molecules presents zig-zag features. These results can be compared to the study of formation of ordered phases of $C_{60}$ (symmetric) as a consequence of the dependence of the diameter of the carbon nanotube [7].

The density of states (DOS) for the CNTs without fullerenes shows an absence of states (gap) near the Fermi energy (EF), thus is a semiconducting as expected. This band gap, for the tight-binding method considered here, is about 0.8, 0.6 and 0.4 eV for (11,0), (13,0) and (17,0) CNTs, respectively. These values are in agreement with results from ab initio calculations [25].

For the $kC_{20}@(n,0)$ structures, our tight-binding study reveals the presence of additional states (associated with the fullerene) close to the conduction band edge [see Fig. 1(a) for $1C_{20}@(11,0)$]. Therefore, the encapsulation of the fullerenes alters the semiconductor characteristics of the empty CNT, shifting the position of the Fermi energy (Fig. 1a), like a conventional $n$-type doping mechanism. The donor state lies about 50 meV below the conduction band and the resulting peapod is expected to behave as an efficient room temperature $n$-doped semiconductor. When the number of fullerenes is increased [$3C_{20}@(11,0)$ and $6C_{20}@(11,0)$, in Fig.1(b) and (c) respectively], more isolated peaks near the conduction band appears. A detailed analysis of fullerene configuration has revealed some disorder where different configuration types can occur. In Figure 2 we show these configurations with the behavior of the potential energy for each of them as a function of the separation distance. The configurations present different dissociation energies as well as equilibrium distance separations. These three local arrangements of fullerenes present in the $6C_{20}@(11,0)$ lead to different interactions with the tube (in contrast to the $C_{30}$ case, see below) and can be associated with the first three isolated peaks presented in the band gap region.

When we considered larger nanotubes to form $kC_{20}@(13,0)$ and $kC_{20}@(17,0)$ peapods (Fig. 3), the states associated with the $C_{20}$ molecules turned out to be located well inside the conduction band, so that they are depleted and their electrons fall to the bottom of the conduction band. Hence, the Fermi level is no longer pinned by the fullerene states and the peapod is metallic. As a consequence, the change of the molecular phase of $C_{20}$ (linear to zig-zag), possible in these larger CNTs, does not significantly affect the characteristics of DOS distributions.

Differently to the case of $kC_{20}@(n,0)$ peapods, the incorporation of $C_{30}$ into the zig-zag nanotubes originates the presence of additional states in the DOS of CNT at the top of the valence band, acting therefore as a $p$-type dopant. The width of these states is proportional to the fullerene concentration, similarly to the case of doping in semiconductors through the incorporation of impurities [26]. This is possible, differently from the case of $C_{20}$ encapsulation discussed above, because $C_{30}$ show a preferential alignment inside the nanotube that leads to a more organized energy level distribution. This can be clearly seen for the $kC_{30}@(11,0)$ peapods (Fig. 4). In this case the position of the Fermi level is in the middle of the band generated by the $C_{30}$ fullerenes.

This situation is altered when larger nanotubes ($n = 13, 17$) are considered (Fig. 5). For the $kC_{30}@(13,0)$ peapods the linear arrangement is still preserved but with fullerenes not so compressed by the tube as in the $kC_{30}@(11,0)$ case. Thus, the whole band associated with the $C_{30}$ fullerenes inside the band gap region is shifted towards the conduction band, also changing the Fermi level position. This behavior also occurs in the $kC_{30}@(17,0)$ case where the $C_{30}$ associated band is even closer to the bottom of the conduction band. However, in this case the larger diameter of nanotube allows more mobility to the fullerenes and some disorder appears in the resulting zigzag arrangement. This causes a distribution of fullerene electronic states similar to the one observed in the $kC_{20}@(n,0)$ peapods (see Fig. 1).

As we have observed n and p type doping, we have also considered the case where both fullerenes types ($C_{20}$ and $C_{30}$) are incorporated into the same nanotube, in order to investigate the possibility of these arrangements work as a pn-juction, envisaging the possibility of realizing a CNT-based nano-diode. For the $k_1C_{20}C_{30}@(11,0)$ and $k_1C_{20}C_{30}@(13,0)$ cases (Fig. 6), we can see that the resulting density of states is a superposition of the cases of the pure encapsulation of $C_{20}$ and $C_{30}$. However, the $C_{20}$ incorporation has the role of pushing the Fermi level towards the conduction band direction. For the encapsulation in larger tubes ($k_1C_{20}C_{30}@(17,0)$) the disorder of the internal fullerenes is even higher and causes a larger spreading of the fullerene states over the energy region close to the Fermi level.

**Conclusions and Discussions**

We have investigated the electronic properties of zigzag carbon nanotubes filled with $C_{20}$ and $C_{30}$ fullerenes (peapods) using a multi-scale approach, involving force field structure determinations and electronic structure from a tight-binding approximation.

The encapsulation of $C_{20}$ and $C_{30}$ fullerenes alters the electronic features of the peapod, which behaves as an *n*- and *p*-type dopant, respectively. The doping is especially efficient in small tubes, *i.e.* (11,0) and (13,0) CNTs. In such a case, the fullerenes provide the nanotube with shallow donor and acceptor levels and the resulting peapod can operate as an efficient room temperature doped semiconductor.

N-type doping is better achieved with a low concentration of $C_{20}$. When the concentration is increased, the linear arrangement of the fullerenes can assume different configurations that give rise to a wider distribution of donor peaks close to the conduction band, some of them being deeper and thus worsening the doping efficiency. On the other hand, $C_{30}$ shows a preferential alignment inside the nanotube and increasing their concentration only results in a broadening of the acceptor level.

For (17,0) nanotubes, a zigzag fullerene pattern begins to emerge. The electronic structure of the zigzag patterns is different from the linear ones due to the larger rotating mobility allowed to the fullerenes inside these encapsulating tubes. This leads to internal fullerene disorder that changes the crystal packing features and consequently the distribution of the electronic levels.

The presence of both fullerene types ($C_{20}$ and $C_{30}$) into the same carbon nanotube (ordered hybrid encapsulation) yields to the superposition of the individual pure behaviors. In such cases, the $C_{20}$ fullerenes show the tendency of pushing the Fermi level towards the conduction band and the $C_{30}$ provides new localized states close to the edge of the valence band, thus giving rise to an effective shrinking of the band-gap. On the other hand, if hybrid encapsulation is ordered, i. e., $C_{20}$s at one side and $C_{30}$s at the other, the peapod could work as a pn-junction, thus allowing the simple fabrication of a CNT-based nano-diode.

In conclusion, the encapsulation of small fullerenes in small diameter semiconducting (zigzag) carbon nanotubes can provide alternative ways for p- and n-doping, generating new possibilities to the band gap tuning necessary in the electronic devices construction using semiconducting carbon nanotubes. We hope the present study stimulates further investigations along these lines.

**Acknowledgements**

The authors acknowledge the financial supported by the Brazilian agencies CNPq, CAPES, and FAPESP, and the computational facilities of the CENAPAD-SP. Support from IMMP/CNPq, IN/CNPq, and SAMNBAS is also acknowledged.

**FIGURE CAPTIONS**

**Figure 1:** Partial density of states (DOS) of the carbon nanotube (black) and the $C_{20}$ fullerenes (red) for the (a) $1C_{20}@(11,0)$, (b) $3C_{20}@(11,0)$, and (c) $6C_{20}@(11,0)$ cases. In all DOS graphs the dashed line indicates the value of Fermi energy and the DOS units are in arbitrary units.

**Figure 2:** The internal fullerene configuration of the $6C_{20}@(11,0)$ peapod is shown in the top and the three different linear arrangements found in it are separately shown in the right (a,b, and c). Results of force field potential energy calculations for each isolated arrangement are shown in the left when the distance between the closest atoms in the arrangement (shown in green in the right) is varied. The equilibrium distances are 2.07, 3.01, and 3.09 Å, for the cases (a), (b), and (c), respectively.

**Figure 3:** Partial density of states (DOS) of the carbon nanotube (black) and the $C_{20}$ fullerenes (red) for the (a) $8C_{20}@(11,0)$, (b) $8C_{20}@(13,0)$, and (c) $8C_{20}@(17,0)$ cases.

**Figure 4:** Partial density of states (DOS) of the carbon nanotube (black) and the $C_{30}$ fullerenes (green) for the (a) $1C_{30}@(11,0)$, (b) $3C_{30}@(11,0)$, and (c) $6C_{30}@(11,0)$ cases.

**Figure 5:** Partial density of states (DOS) of the carbon nanotube (black) and the $C_{30}$ fullerenes (green) for the (a) $4C_{30}@(11,0)$, (b) $4C_{30}@(13,0)$, and (c) $4C_{30}@(17,0)$ cases.

**Figure 6:** Partial density of states (DOS) of the carbon nanotube (black), $C_{20}$ (red), and the $C_{30}$ fullerenes (green) for the (a) $4C_{20}C_{30}@(11,0)$, (b) $4C_{20}C_{30}@(13,0)$, and (c) $4C_{20}C_{30}@(17,0)$ cases.

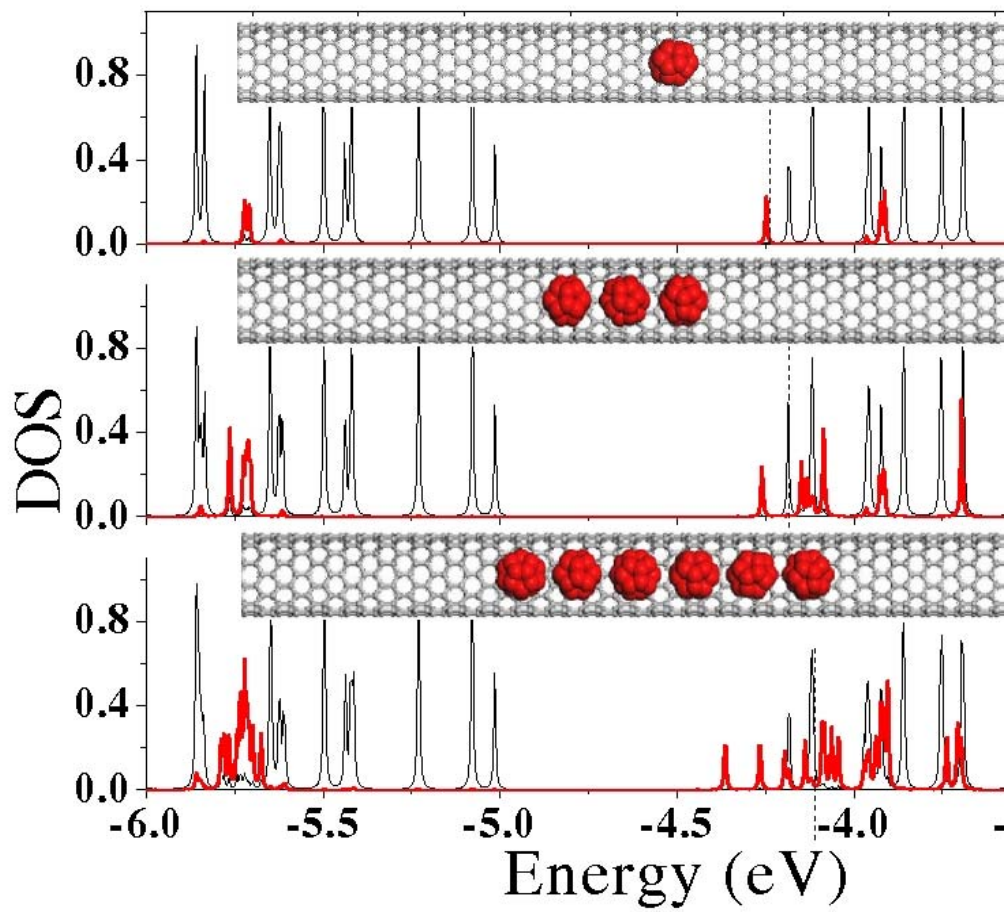

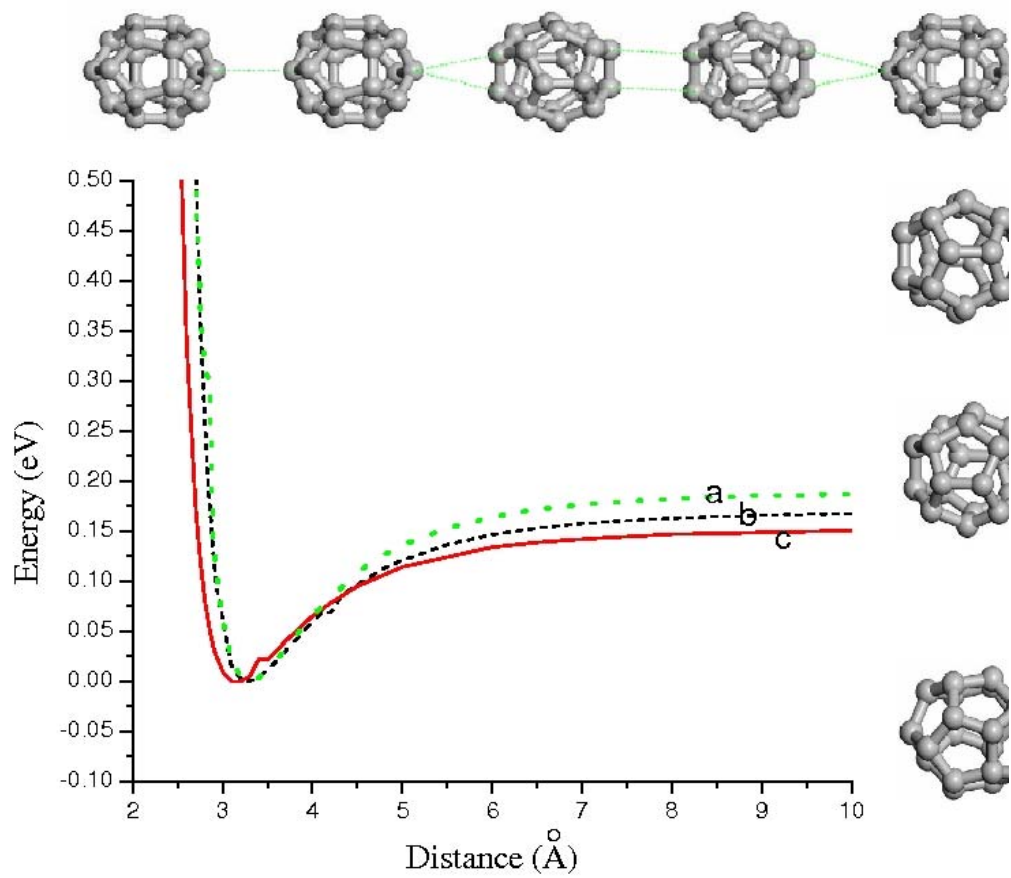

FIG. 2

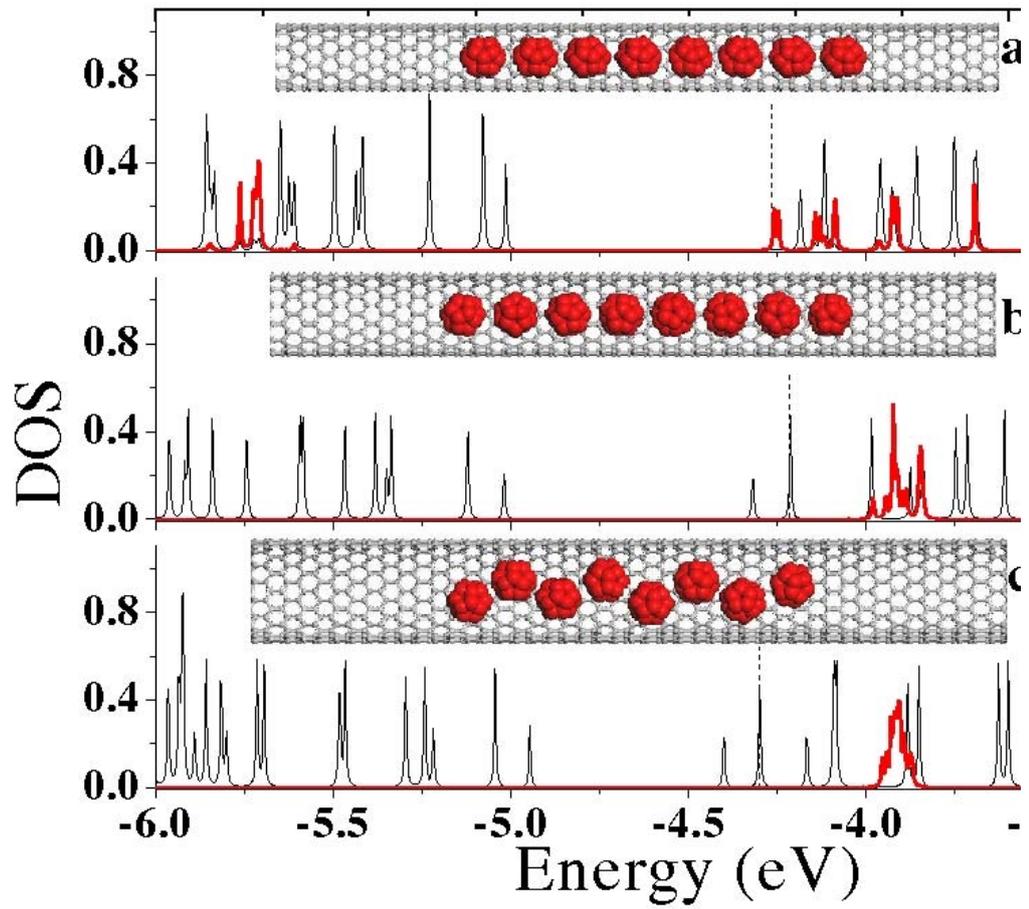

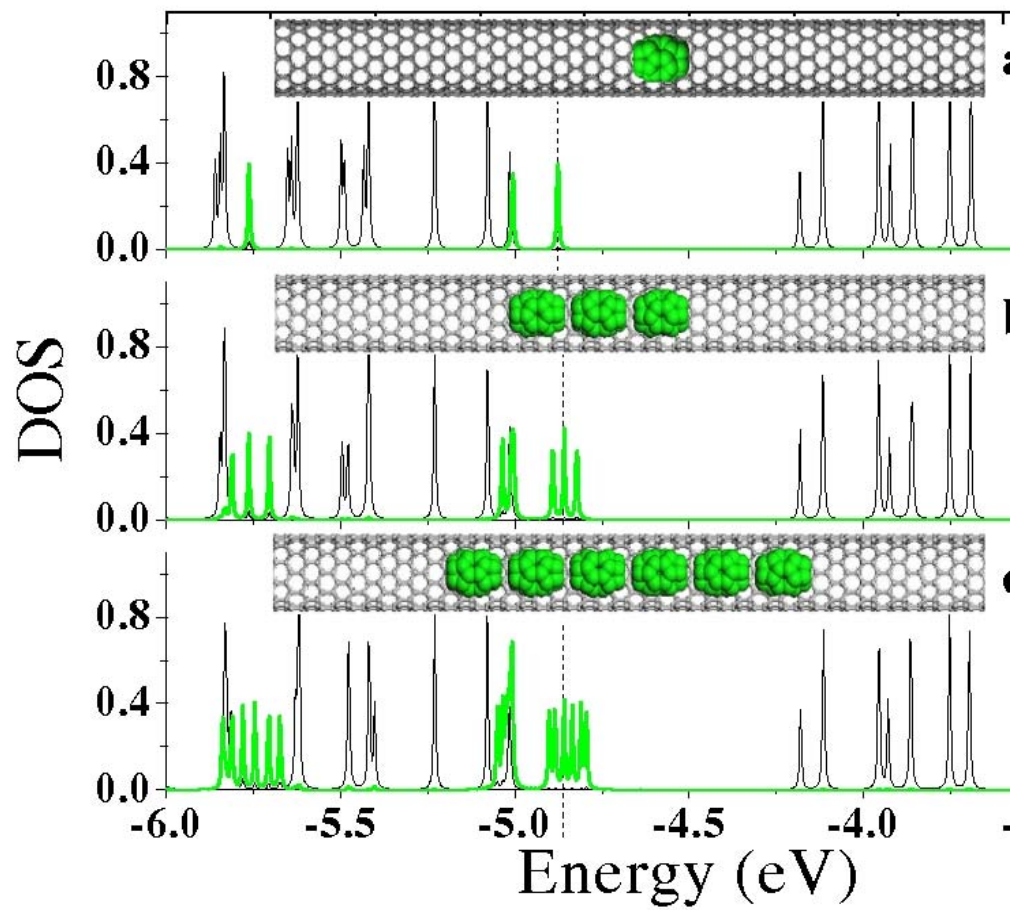

FIG. 4

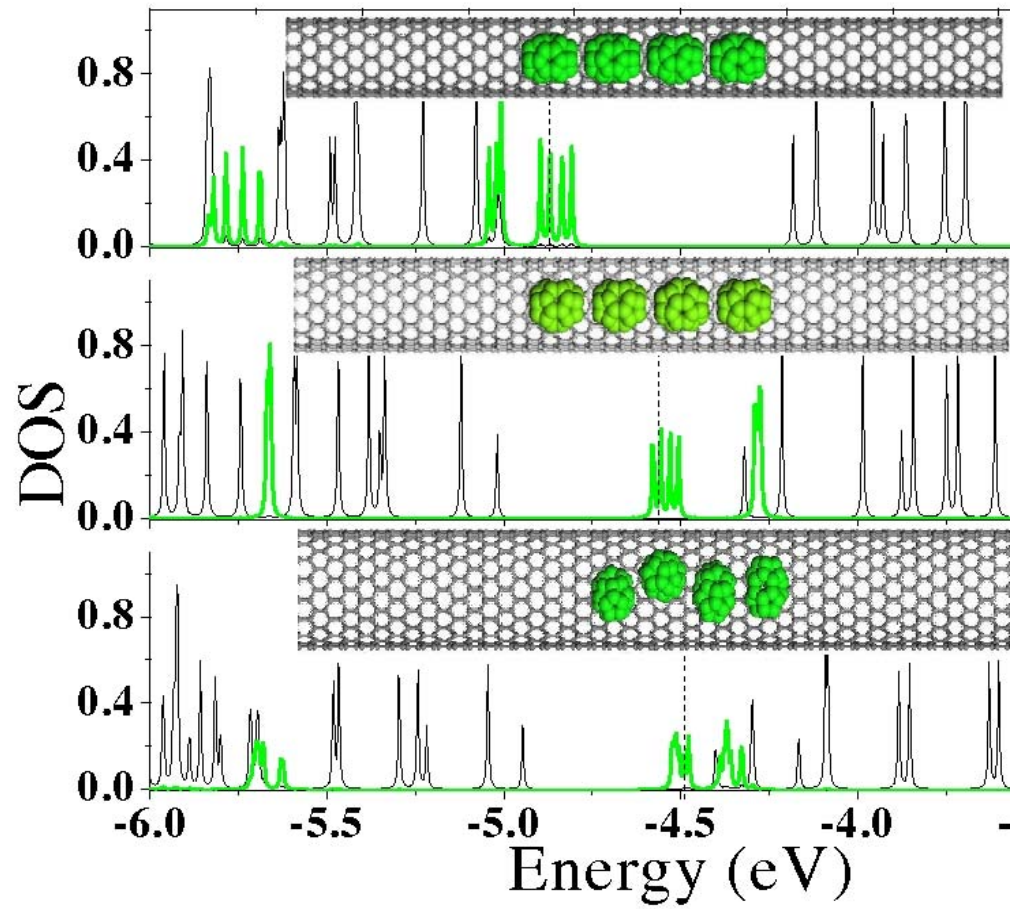

FIG. 5

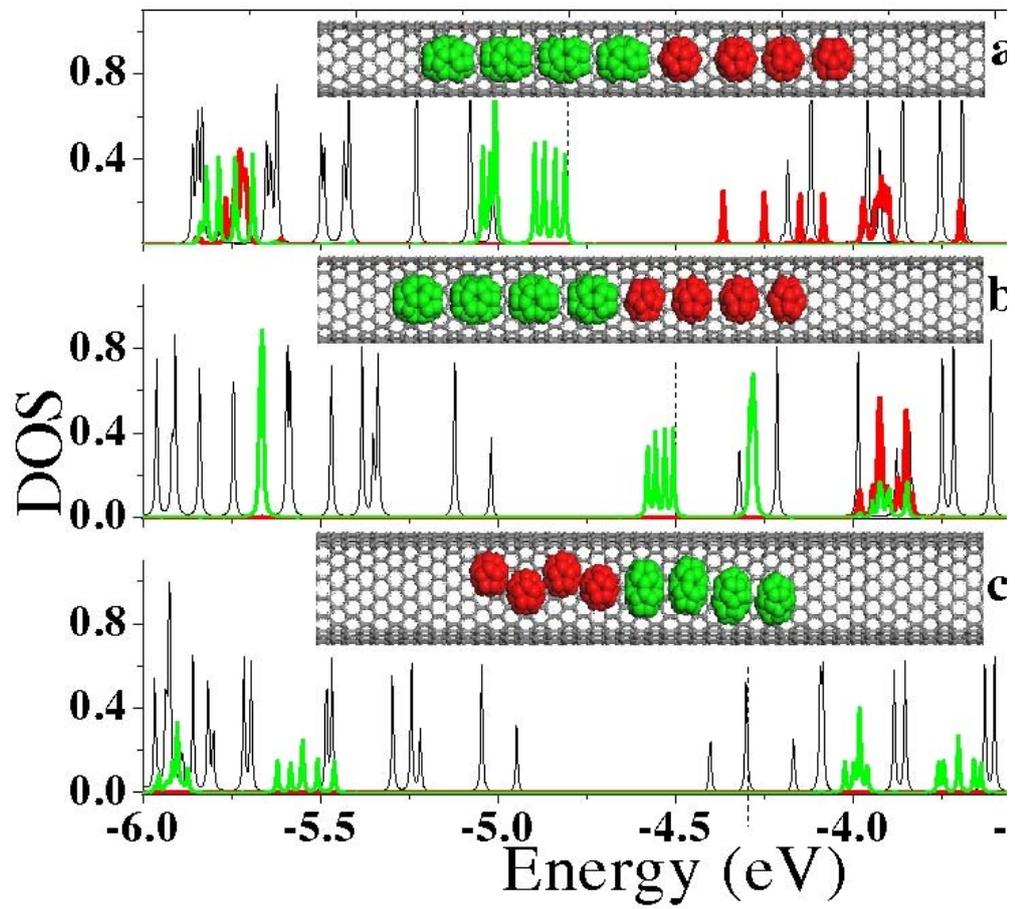

FIG.6